# Electronic structure and bonding properties of cobalt oxide in the spinel structure

*Jia Chen[1], Xifan Wu[2], Annabella Selloni[1,\*]*

[1]Department of Chemistry, Princeton University, Princeton, New Jersey 08544, USA

[2]Department of Physics, Temple Materials Institute,and Institute for Computational Molecular Science,Temple University, Philadelphia, PA 19122, USA

April 12, 2011

ABSTRACT. The spinel cobalt oxide $Co_3O_4$ is a magnetic semiconductor containing cobalt ions in $Co^{2+}$ and $Co^{3+}$ oxidation states. We have studied the electronic, magnetic and bonding properties of $Co_3O_4$ using density functional theory (DFT) at the Generalized Gradient Approximation (GGA), GGA+U, and PBE0 hybrid functional levels. The GGA correctly predicts $Co_3O_4$ to be a semiconductor, but severely underestimates the band gap. The GGA+U band gap (1.96 eV) agrees well with the available experimental value (~ 1.6 eV), whereas the band gap obtained using the PBE0 hybrid functional (3.42 eV) is strongly overestimated. All the employed exchange-correlation functionals predict 3 unpaired d electrons on the $Co^{2+}$ ions, in agreement with crystal field theory, but the values of the magnetic moments given by GGA+U and PBE0 are in closer agreement with the experiment than the GGA value, indicating a better description of the cobalt localized d states. Bonding properties are studied by means of Maximally Localized Wannier Functions (MLWFs). We find d-type MLWFs on the cobalt ions, as well as Wannier functions with the character of $sp^3d$ bonds between cobalt and oxygen ions. Such hybridized bonding states indicate the presence of a small covalent component in the primarily ionic bonding mechanism of this compound.

## 1. Introduction

In the current search for new efficient catalysts for water oxidation, tricobalt tetraoxide, $Co_3O_4$, has emerged as a particularly promising material for various applications in energy and environment-related areas.[1-4] In particular, significant progress toward artificial photosynthetic systems has been achieved using nanostructured $Co_3O_4$ to catalyze water oxidation under mild conditions.[1] $Co_3O_4$ is also an efficient catalyst for methane combustion[2] and CO oxidation at low temperatures,[4] and has been used as the anode material of lithium-ion batteries and as a gas sensor.[3]

Despite these important technological applications, the amount of available information on $Co_3O_4$ is still limited. $Co_3O_4$ crystallizes in the cubic normal spinel structure (space group $Fd\bar{3}m$) which contains cobalt ions in two different oxidation states, $Co^{2+}$ and $Co^{3+}$. These are located at the interstitial tetrahedral (8$a$) and octahedral (16$d$) sites, respectively, of the close-packed face centered cubic (fcc) lattice formed by the oxygen ions (see Fig. 1). In a simplified picture, the crystal fields at the 8$a$ and 16$d$ sites split the five degenerate atomic d orbitals into two groups, leading to 3 unpaired d electrons on $Co^{2+}$, while all the d electrons of $Co^{3+}$ are paired (see Fig. 2). As a result, the $Co^{3+}$ ions are not magnetic, whereas the $Co^{2+}$ ions carry a magnetic moment.

Experimentally, $Co_3O_4$ is a paramagnetic semiconductor at room temperature. It becomes antiferromagnetic below $T_N \sim 40$ K,[5] where the antiferromagnetism is mainly due to the weak coupling between nearest neighbor $Co^{2+}$ ions. The conductivity is usually p-type at low temperature and intrinsic at high temperature;[6] measured values of the band gap are around 1.6 eV.[7,8]

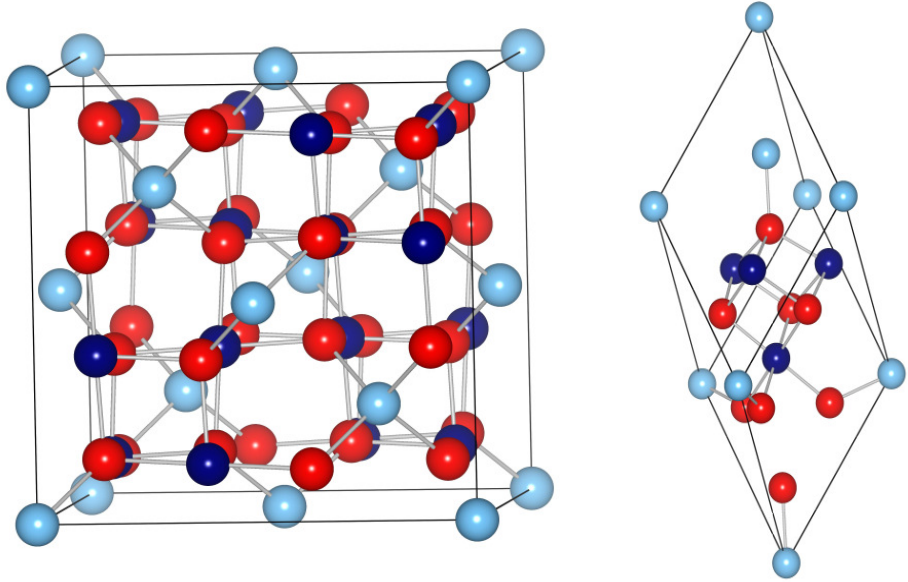

*FIG. 1. Unit cell (on the left) and primitive cell (on the right) of $Co_3O_4$. Light cyan and navy blue balls indicate $Co^{2+}$ and $Co^{3+}$ ions, red ones indicate $O^{2-}$ ions.*

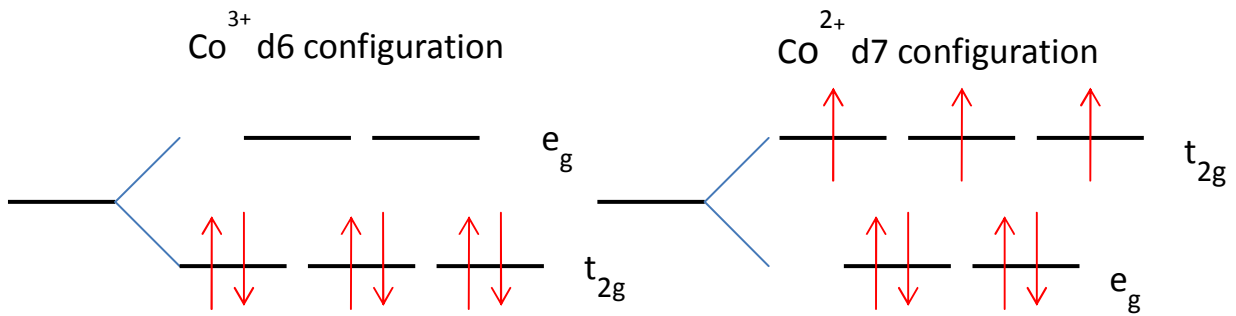

*FIG. 2. Schematic diagram of crystal field splitting of $Co^{3+}$ ion in octahedral field (on the left) and $Co^{2+}$ ion in tetrahedral field (on the right).*

First principles theoretical studies of the properties of $Co_3O_4$ are not numerous,[9,10] and the bonding mechanism and electronic structure of this material have not been fully elucidated yet. The main objective of the present work is to fill this gap, and provide a comprehensive theoretical analysis of the structural, electronic, magnetic, and bonding properties of $Co_3O_4$ based on first principles electronic structure methods. As a transition metal oxide semiconductor with a complex atomic and magnetic structure, $Co_3O_4$ is quite challenging to describe using ab-initio methods. Standard density functional theory (DFT) calculations within the local density approximation (LDA) or the generalized gradient approximation (GGA) often severely underestimate the band gap due to the delocalization error arising from the incomplete cancellation of the Coulomb self-interaction.[11] Among the schemes proposed to overcome this problem, the DFT+U method[12] and hybrid functionals[13] have been applied with some success to a number of solid-state materials.[14] In this work, we investigate the bulk properties of $Co_3O_4$ using both these approaches in conjunction with standard DFT-GGA. We determine the U values for the GGA+U calculations from first principles, using linear response.[15] For the hybrid functional calculations, we adopt the "parameter-free" PBE0 functional[16] and use a recently introduced order-N implementation[17] based on Maximally localized Wannier functions (MLWFs).[18] MLWFs are also used also to analyze the bonding properties and the character of the localized d states, and the MLWFs from the different electronic structure approaches are compared.

Following a short description of the computational details in Sec. II, our results for the electronic, bonding and magnetic properties of $Co_3O_4$ are presented and discussed in Sec. III. Summary and conclusions are given in Sec. IV.

## 2. Methods and computational details

The DFT-GGA and GGA+U calculations were performed using the plane wave-pseudopotential scheme as implemented within the Quantum Espresso package.[19] All calculations were spin-polarized and the exchange-correlation terms were described using the Perdew-Burke-Ernzerhof (PBE) functional.[20] Norm-conserving Troullier-Martins pseudopotential[21,22] were employed; the Co(3d, 4s) pseudopotential included nonlinear core corrections.[23] A plane-wave kinetic energy cut-off of 120 Ry was chosen, which ensures a good convergence of the computed lattice constant (see Table I). Calculations were performed on the 14-atom primitive unit cell of the spinel structure. An 8×8×8 k-point grid was used to obtain a well converged sampling of the Brillouin zone. [24]

**Table I**. Lattice constant (Å) computed at the GGA-PBE level as a function of the plane wave kinetic energy cutoff Ecut (in Ry).

| Ecut | 70 | 100 | 120 | 150 |
|---|---|---|---|---|
| **Lattice constant** | 8.01 | 8.15 | 8.19 | 8.20 |

_Calculation of U._ We determined the Hubbard U parameter for the $Co^{2+}$ and $Co^{3+}$ ions of $Co_3O_4$ using the linear response approach of Ref.15. In order to avoid possible interference effects caused by the periodic boundary conditions, calculations were performed on various supercells with volumes ranging from one to four primitive cells. Converged values of the effective U parameter (corresponding to U-J in the original LDA+U formulation)[15] are 4.4 and 6.7 eV for $Co^{2+}$ and $Co^{3+}$, respectively. Our computed U value for $Co^{2+}$ is very close to that obtained for CoO in previous work.[25] The fact that the value of U is larger for $Co^{3+}$ than for $Co^{2+}$ is due to the stronger on-site repulsion in the more contracted d orbitals of ions with higher oxidation state. Unless otherwise specified, the above U values for $Co^{2+}$ and $Co^{3+}$ are used in all the GGA+U (also referred to as PBE+U) calculations of this work.

*PBE0 calculations* Hybrid functional PBE0[16] calculations of the electronic and magnetic properties were performed using the approach of Wu et al.[17] The PBE0 hybrid functional is constructed by mixing 25% of exact exchange ($E_x$) with the GGA-PBE exchange ($E_{PBEx}$), while the correlation potential is still represented by the corresponding functional in PBE ($E_{PBEc}$)

$$E_{PBE0xc} = (1/4)E_x + (3/4)E_{PBEx} + E_{PBEc} .$$

$E_x$ has the usual Hartree-Fock form in terms of one-electron orbitals. In the method of Ref. [17] this term is expressed in terms of localized Wannier orbitals. These are obtained through an unitary transformation of the delocalized Bloch states corresponding to occupied bands (see Sect. 3.3). In particular, we use maximally localized Wannier functions (MLWFs)[18], which are exponentially localized. In this way, a significant truncation in both number and size of exchange pairs can be achieved in real space. The cutoff value for the truncation of pair-exchange energies was set at $10^{-5}$ a.u.[17] Both the conventional cubic cell with 56 atoms and a tetragonal supercell containing 112 atoms (corresponding to twice the conventional cell) were considered, with k-sampling always restricted to the Γ point. Test calculations performed at the GGA-PBE and PBE+U levels showed the consistency of the results obtained with this setup with those obtained using the primitive unit cell and an 8×8×8 k-space mesh, the band gap and band width differences between the two setups being 0.03(0.02) and 0.21 (0.02) eV, respectively, within PBE (PBE+U), see Section 3.2.

## 3. Results and discussion

### 3.1 Structural properties

Our results for the structural properties of $Co_3O_4$ are summarized in Table II. The lattice constant was determined by fitting computed total energies to Murnaghan's equation of state,[26] with all internal degrees of freedom fully relaxed. Comparison to experiment shows that the GGA-PBE lattice constant and bond distances are overestimated by about 1.5 percent. With PBE+U, the overestimate increases to about 2 percent, as found also in other GGA+U studies of oxide materials.[27,28]

**TABLE II.** Lattice constant (Å), bulk modulus (GPa), and bond distances (Å) of $Co_3O_4$ from PBE and PBE+U calculations using the primitive 14-atom unit cell and an 8×8×8 k-point mesh.

|  | PBE | PBE+U | Expt. |
|---|---|---|---|
| **Lattice constant** | 8.19 | 8.27 | 8.08 |
| **Bulk Modulus** | 199 | 192 | --- |
| **Distance $Co^{2+}$-$O^{2-}$** | 1.95 | 1.99 | 1.93 |
| **Distance $Co^{3+}$-$O^{2-}$** | 1.93 | 1.95 | 1.92 |

For the fully optimized structure, we determined the heat of formation ($\Delta H$) at T=298 K relative to metallic, ferromagnetic Co in the hcp structure and gas phase $O_2$. For Co, the PBE+U reference was obtained by combining the ground state energies at U=4.4 and 6.7eV in a 1:2 ratio;

Co was computed to be metallic and ferromagnetic at both values of U. We found $\Delta H$ = -683 and -815 kJ/mol at the PBE and PBE+U levels, respectively, whereas experimentally values of -891[29] and -910[30] kJ/mol have been reported. To avoid the difficulties associated with having two different U values for Co, we also performed calculations for $Co_3O_4$ using a single value of U for both $Co^{2+}$ and $Co^{3+}$, namely U = 4.4, 5.9 and 6.7 eV, the value U= 5.9 eV being a 1:2 average of 4.4. and 6.7 eV. From these calculations, we obtained $\Delta H$= -803, -852, and -884 kJ/mol, respectively, indicating that the dependence of the computed $\Delta H$ on the value of U is moderate, a 10% variation for a variation of U of more than 2eV. Comparison between the value of $\Delta H$ obtained using the average U value, U=5.9 eV, and experiment shows a deviation of the order 6%.

It has been suggested that the difference between theoretical and experimental values of $\Delta H$ may be often attributed to the overestimate of the $O_2$ binding energy ($E_b$) given by GGA calculations.[31] Using the experimental value of $E_b$, we find $\Delta H_{corr}$ = -878 kJ/mol at the PBE level, in reasonable agreement with the experiment. Instead, using PBE+U with U=5.9 eV for both $Co^{2+}$ and $Co^{3+}$, we obtain $\Delta H_{corr}$ = -1047 kJ/mol, which is significantly overestimated in comparison to experiment.

### 3.2 Electronic properties

Results of PBE and PBE+U calculations of the electronic band structure along various symmetry directions in the Brillouin zone are presented in Fig. 3. Computed direct and indirect gaps at a few symmetry points are reported in Table III. Both PBE and PBE+U predict the valence band maximum and the conduction band minimum to occur at the high symmetry point X along the [100] direction, thus resulting in a direct minimum band gap at X. The GGA-PBE approach successfully predicts $Co_3O_4$ to be a semiconductor but the minimum band gap, 0.3 eV, is severely underestimated with respect to the experimental value of 1.6 eV (obtained from measurements on films and nanocrystalline samples).[7,8] The PBE+U method gives a minimum gap of 1.96eV, in satisfactory agreement with the experiment. To test the sentivity of the gap to the use of two different U values for $Co^{2+}$ and $Co^{3+}$, we also performed calculations with a single U for both $Co^{2+}$ and $Co^{3+}$. We found a gap of 1.67, 2.02, and 2.16 eV using U = 4.4, 5.9 and 6.7 eV, respectively. The band gap with the average U=5.9 eV value is thus very similar to that obtained using two different values of U for the $Co^{2+}$ and $Co^{3+}$ ions. Both PBE and PBE+U approaches predict a larger dispersion near the conduction band minimum than at the valence band maximum, and therefore a smaller effective mass for electron than for hole states.

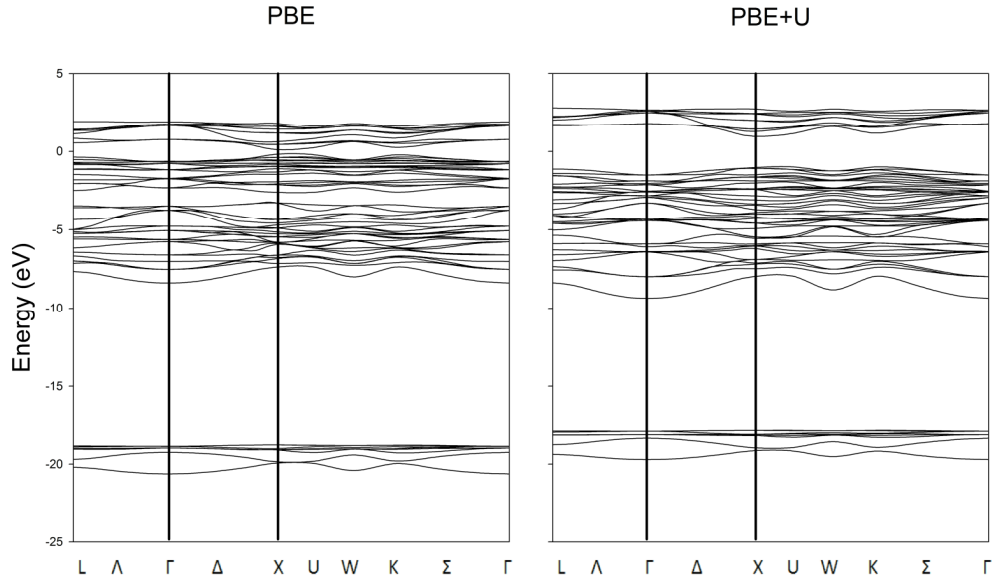

*FIG. 3. Band structure of $Co_3O_4$ obtained by PBE (left) and PBE+U (right) approach. Fermi energy is set to 0.*

Figure 4 displays the GGA-PBE and PBE+U partial densities of states (PDOS) obtained by projecting the Kohn-Sham states onto atomic orbitals centered on the various cobalt and oxygen ions. The GGA-PBE results show a splitting of the valence band in two sub-bands. The sub-band at lower energies is dominated by O 2p states, while the upper one originates primarily from $Co^{3+}$ d states, especially in proximity of the valence band edge. In the upper valence band, smaller contributions from oxygen states and $Co^{2+}$ d states are also present, with a $Co^{2+}$ peak around -2.5 eV. These features are in qualitative agreement with results of photoemission experiment performed on a $Co_3O_4$ film epitaxially grown on CoO(100).[32] The bottom of the conduction band shows contributions of both $Co^{2+}$ and $Co^{3+}$ d states, with a similar weight.

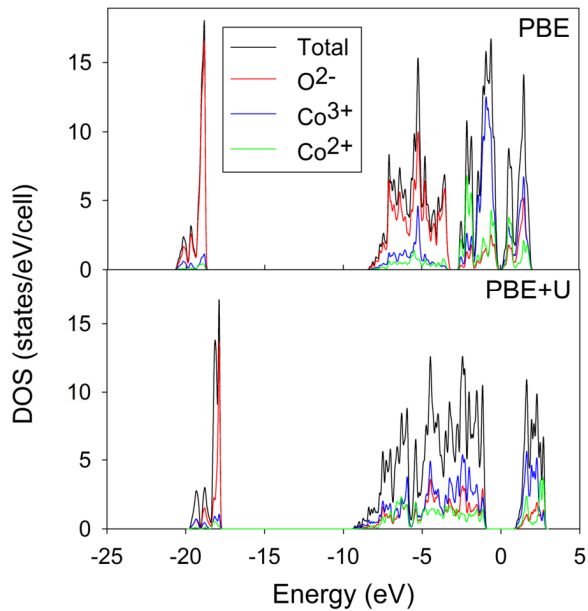

*FIG. 4. Total and projected density of states from PBE (top) and PBE+U (bottom) calculations.*

*Fermi energy is set to 0.*

In the PBE+U results, the total valence bandwidth, 8.33 eV, is similar to that given by GGA-PBE, 8.20 eV. At variance with the pure PBE case, however, no clear splitting of the valence band is present. The contributions from O 2p, $Co^{3+}$ d and $Co^{2+}$ d states in the PDOS are spread with similar weights throughout the valence band, indicating a stronger hybridization with respect to the PBE case. This can be attributed to a stabilization of the cobalt d orbitals relative to the O 2p and 2s states. The band gap is much wider than in GGA-PBE, and at the bottom of the conduction band the largest contribution originates from the $Co^{3+}$ d states.

**TABLE III.** Direct and indirect band gaps (in eV) at various symmetry k-points, from PBE and PBE+U calculations. As a comparison, the minimum gap obtained from PBE0 calculations is 3.42 eV, while experiments found 1.60 (Ref.7) and 1.65 eV (Ref. 8).

| Band gap | PBE | PBE+U |
|:---:|:---:|:---:|
| Γ->Γ | 1.39 | 3.25 |
| X->Γ | 0.94 | 2.81 |
| Γ->X | 0.75 | 2.41 |
| X->X | 0.30 | 1.96 |

Calculations based on the PBE0 hybrid functional were performed at the Γ point of a tetragonal supercell containing 112 atoms, which includes the X point of the primitive cell. The experimental lattice constant and geometry were employed. For a more direct comparison, calculations using the same setup were performed also at the PBE and PBE+U levels. Both the valence bandwidth (9.48 eV) and the band gap (3.42 eV) obtained with the PBE0 functional are larger than those given by PBE (8.41 and 0.33 eV, respectively) and PBE+U (8.35 and 1.94 eV), a trend observed for other oxide semiconductors as well, see, e.g., Refs. [33] and [34]. This trend, however, appears to be amplified in the present case, resulting in a substantial overestimate of the computed band gap with respect to the experiment.

### 3.3 Bonding properties

Numerous studies have shown the effectiveness of maximally-localized Wannier functions (MLWFs) for the analysis of the electronic and bonding properties of crystalline materials.[35,36] In particular, the Wannier centers and the shapes of MLWFs have been found to provide useful insights into the nature of chemical bonds in a variety of compounds. To investigate the bonding properties of $Co_3O_4$, we determined the MLWFs at the PBE, PBE+U and PBE0 levels, using the algorithm developed by Sharma et al.[37] We considered the conventional 56-atom cubic supercell and restricted k-space integration to Γ only. Our results are summarized in Table IV.

Independent of the approach used to describe the electronic structure, we found 6 and 7 singly occupied d-type Wannier functions whose centers are very close to each cobalt ion at an octahedral and tetrahedral site, respectively. Since with our pseudopotential a neutral Co should have 9 valence electrons, this means that the charge states of the cobalt ions at octahedral and tetrahedral sites are $Co^{3+}$ and $Co^{2+}$, respectively, in full agreement with their expected oxidation states. Similarly, we found 4 pairs of Wannier centers (WCs) in proximity of each oxygen ion (see Figure 5), indicating an $O^{2-}$ charge state, in agreement with the formal oxidation state of oxygen ions. The simple connection between Wannier centers and ionic oxidation states is quite remarkable. Standard computational approaches for characterizing oxidation states are based on electron population analyses, in which the occupied electronic states are projected onto atom-centered orbitals. It is known, however, that these methods depend significantly on the basis set chosen and often yield charges that cannot be directly related to the oxidation states.[38-40] In fact, if we integrate the occupied PDOS for $Co^{3+}$ and $Co^{2+}$ (e.g. the DOS computed at the PBE level in Fig.4), we find more electrons on $Co^{3+}$ (7.98) than on $Co^{2+}$ (7.70), in obvious contrast with the expected oxidation states, simply because the $Co^{3+}$ ions are coordinated to a larger number of oxygens, and therefore more electrons from Co-O bonds are around them. This clearly shows that projections onto atomic-like states are inadequate for the identification of oxidation states. It was recently pointed out that optimally-localized orbitals, such as MLWFs in the case of crystalline systems, can instead provide a reliable estimate of oxidation states.[41] This is confirmed by our results for $Co_3O_4$, for which the analysis of Wannier centers unambiguously identifies cobalt ions in +2 and +3 oxidation states.

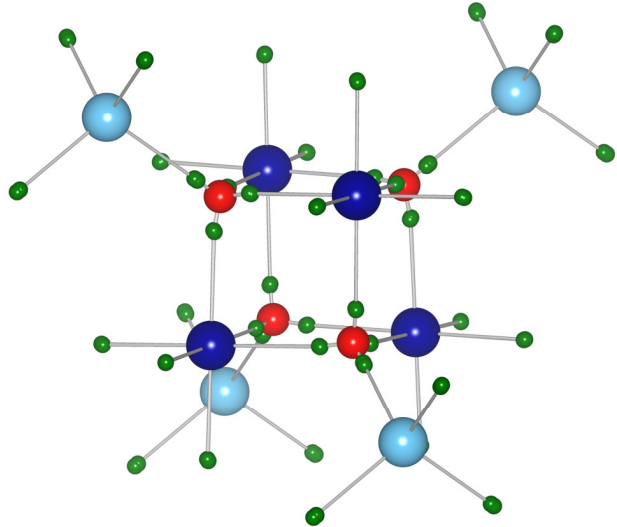

FIG. 5. Wannier Centers of $Co_3O_4$. Light cyan, navy blue and red balls indicate $Co^{2+}$, $Co^{3+}$ and $O^{2-}$ ions respectively. Green small balls indicate Wannier Centers near $O^{2-}$ ions. Wannier centers very close to Co ions are almost overlap with Co ions so that can not been seen in this figure.

**TABLE IV**. Type and number of Wannier functions for each ion, spreads $\Omega$ of MLWFs (in $a_0^2$), average energy $E_{MLWF}$ of MLWFs (in eV) relative to the energy of $Co^{2+}$ $t_{2g}$. Results of PBE, PBE+U and PBE0 calculations are listed.

| MLWF type | No. | PBE | | PBE+U | | PBE0 | |
|---|---|---|---|---|---|---|---|
| | | $\Omega$ | $E_{MLWF}$ | $\Omega$ | $E_{MLWF}$ | $\Omega$ | $E_{MLWF}$ |
| $Co^{3+}$ $t_{2g}$ | 6 | 0.66 | 1.1 | 0.47 | 2.2 | 0.46 | 0.6 |
| $Co^{2+}$ $t_{2g}$ majority Spin | 3 | 0.76 | **0.0** | 0.51 | **0.0** | 0.54 | **0.0** |
| $Co^{2+}$ $e_g$ majority Spin | 2 | 0.53 | 0.6 | 0.45 | 1.1 | 0.45 | 0.8 |
| $Co^{2+}$ $e_g$ minority Spin | 2 | 0.71 | 1.9 | 0.51 | 2.3 | 0.53 | 2.6 |
| $Co^{3+}$-O $sp^3d$ | 6 | 0.66 | -5.9 | 0.64 | -4.7 | 0.59 | -5.6 |
| $Co^{2+}$-O $sp^3d$ | 2 | 0.64 | -6.8 | 0.61 | -5.4 | 0.57 | -6.2 |

The Wannier centers of $Co_3O_4$ are shown in Figure 5. The oxygen WCs are located along the directions connecting each oxygen ion to its four nearest neighbors, so as to form a somewhat distorted tetrahedron, suggesting that the bonding has a partially covalent character. This is confirmed by the explicit form of the Wannier functions. From Table IV and Figure 6, we can see that the Wannier functions can be classified in 6 different types. These include d states of $t_{2g}$ symmetry localized on $Co^{3+}$ and $Co^{2+}$ ions, d $e_g$ states for majority and minority spins on $Co^{2+}$ ions, and Wannier functions with the character of $sp^3d$ bonds both between $Co^{3+}$ and $O^{2-}$ and between $Co^{2+}$ and $O^{2-}$ ions. These MLWFs show that the bonding character of $Co_3O_4$, although mainly ionic, has also a small covalent component. This is in agreement with earlier work indicating that covalent bonds are essential to cation ordering in the spinel structure.[42]

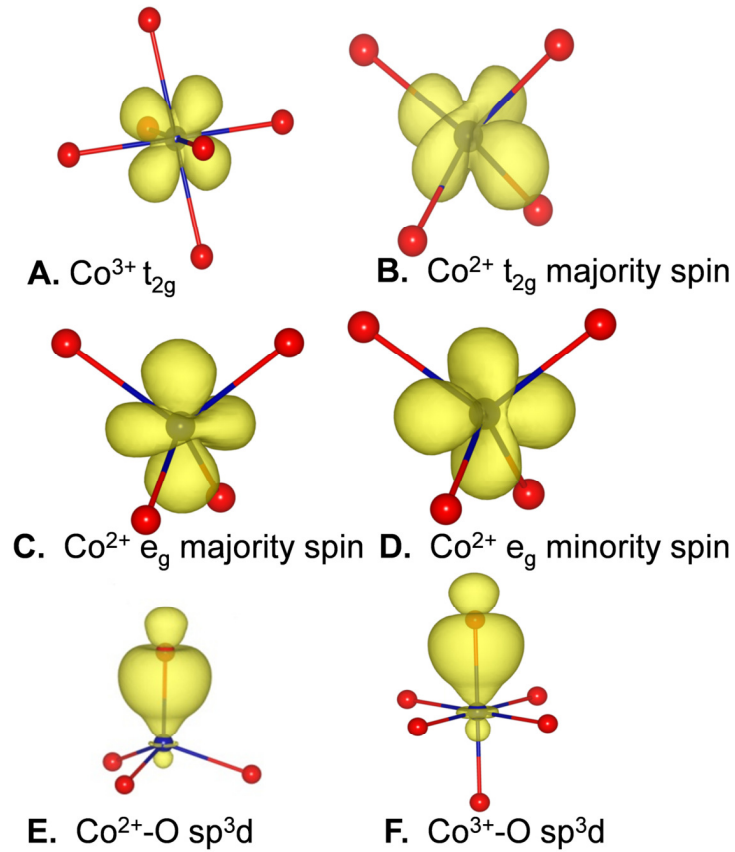

*FIG. 6. Isosurfeces of charge density of 6 types of MLWFs. Value of isosurfaces is 1% of maximal value. Co and O ions are denoted by blue and red balls respectively.*

It is interesting to examine the MLWFs' spreads in Table IV. The d-type and $sp^3d$ bonding MLWFs have similar spreads at the PBE level, whereas the PBE+U and PBE0 calculations predict that the spreads of the d-type MLWFs are smaller than those of the $sp^3d$ bonding orbitals. For the latter, the spreads obtained from PBE and PBE+U calculations are rather similar, whereas those given by PBE0 are smaller. Altogether, PBE0 leads to a stronger localization of all orbitals relative to PBE, whereas the main differences between PBE+U and PBE concern the localization of the d orbitals. We also notice that at all electronic structure levels the average spread of the $Co^{2+}$ d orbitals is larger than the d orbital spread for the $Co^{3+}$ ions. This is consistent with the larger value of U found for $Co^{3+}$ in comparison to that for $Co^{2+}$ (see Sect. II).

In Table IV we also report the average energies $E_{MLWF}$ of the various MLWFs relative to the energy of $Co^{2+}$ $t_{2g}$. $E_{MLWF}$ is defined as

$$E_{MLWF}^i = \sum_j \left| A_{ij} \right|^2 E_{KS}^j \quad (1)$$

where A is the unitary transformation matrix between MLWFs, $\Psi_{MLWF}$, and Kohn-Sham states, $\Psi_{KS}$

$$\psi_{MLWF}^i = \sum_j A_{ij} \psi_{KS}^j \qquad (2)$$

As shown in Fig. 2, crystal field theory predicts doubly occupied $e_g$ orbitals for $Co^{2+}$ ions. Instead, we find that the $Co^{2+}$ $e_g$ orbitals are split into different spin-orbitals due to the exchange interaction between $e_g$ and singly occupied $t_{2g}$ states (see Fig. 7). PBE0 gives the largest splitting, 2.2eV, whereas PBE and PBE+U give splittings of 1.3 and 1.2 eV, respectively. Also different from crystal field theory, the $Co^{2+}$ $e_g$ orbitals are higher in energy than the $Co^{2+}$ $t_{2g}$ orbitals. For the minority $e_g$ spin-orbitals, the higher energy can be attributed to the lack of exchange interaction with the $t_{2g}$ orbitals. For the majority $e_g$ spin-orbitals, the higher energy is likely caused by the Hartree repulsion with the minority spin $e_g$ orbitals. As shown in Fig. 6, there is indeed a strong overlap between majority and minority spin $e_g$ orbitals. Thus the main contribution to the $Co^{2+}$ atomic magnetic moment comes from the three singly occupied $t_{2g}$ orbitals and the resulting moment is close to three.

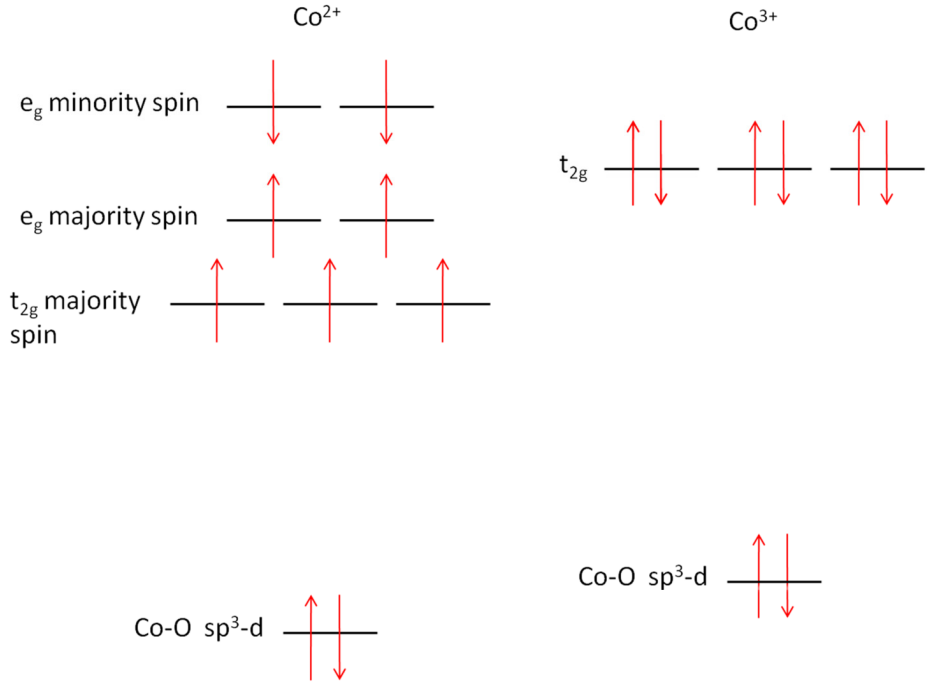

*FIG. 7. Schematic diagram of relative energies of 6 types MLWFs*

### 3.4 Magnetic properties

The magnetic properties of $Co_3O_4$ originate from the atomic magnetic moments of $Co^{2+}$ ions ($\mu_{Co2+}$). Experimentally $\mu_{Co2+} = 3.26\mu_B$, where the part in excess of 3 is due to the spin-orbit coupling.[5] Computed values of $\mu_{Co2+}$ are listed in . PBE+U and PBE0 predict larger values of $\mu_{Co2+}$ with respect to PBE, in better agreement with the experiment. Since the $Co^{2+}$ magnetic moment is associated to localized d electrons, this improvement comes from the partial correction of the PBE delocalization error within PBE+U and PBE0.

Couplings between atomic magnetic moments are weak in $Co_3O_4$, as implied by its low Neel temperature, and may not have a strong impact on the material properties at room temperature. We estimated the coupling parameter $J_1$ between nearest neighbor $Co^{2+}$ ions from the total energy difference between the anti-ferromagnetic and ferromagnetic solutions in the primitive cell using the Heisenberg spin Hamiltonian

$$H = -\sum_{ij} J_1 \vec{S}_i \cdot \vec{S}_j \qquad (3)$$

where i and site j denote nearest neighbor sites. There are 4 nearest neighbor pairs in a primitive cell, and therefore $J_1$ can be expressed as

$$J_1 = \frac{1}{2} \cdot \frac{1}{4} \cdot \frac{1}{S^2} (E_{AFM} - E_{FM}) \qquad (4)$$

where $S = \frac{3}{2}$. Computed values of $J_1$ are reported in Table V. PBE and PBE0 calculations correctly predict the antiferromagnetic phase to be more stable than the ferromagnetic one; however the absolute value of the computed $J_1$ is much larger than the experimental one. By contrast, we found that PBE+U favors the ferromagnetic solution; therefore the computed $J_1$ has opposite sign with respect to the experiment.

**TABLE V.** Atomic magnetic moment of $Co^{2+}$, $\mu_{Co2+}$ (in $\mu_B$), and exchange coupling between nearest neighbors, $J_1$ (in eV), as given by PBE, PBE+U and PBE0 calculations.

| | PBE | PBE+U | PBE0 | Expt. |
|---|---|---|---|---|
| $\mathbf{\mu_{Co2+}}$ | 2.64 | 2.84 | 2.90 | 3.26 (Ref. [5]) |
| $\mathbf{J_1}$ | $-2.5\times10^{-3}$ | $1.0\times10^{-4}$ | $-5.0\times10^{-3}$ | $-6.26\times10^{-4}$ (Ref. [43]) |

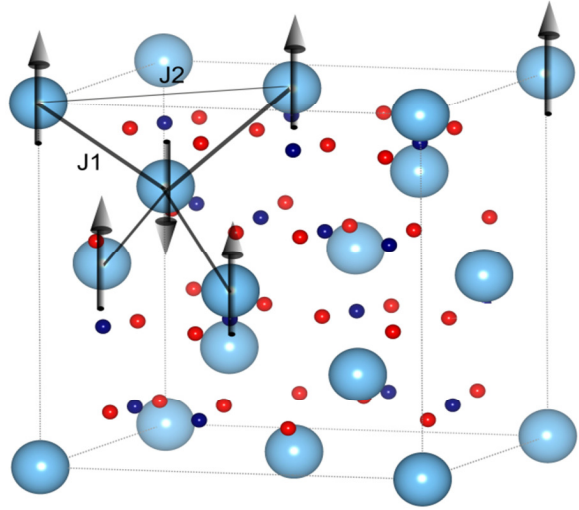

*FIG. 8. Antiferromagnetic configuration and coupling coefficient of $Co_3O_4$. J1 corresponds to magnetic coupling between nearest $Co^{2+}$ neighbors and J2 corresponds to second nearest neighbors.*

## 4. Summary and conclusions

In this work, we have studied the bulk properties of the spinel cobalt oxide $Co_3O_4$ using density functional theory at the PBE, PBE+U and PBE0 levels. The U parameters for the PBE+U calculations were determined from first principles,[15] resulting in 4.4 and 6.7 eV for $Co^{2+}$ and $Co^{3+}$, respectively. The PBE0 calculations were carried out using an order-N method based on Maximally Localized Wannier Functions (MLWFs).[17] The GGA-PBE correctly predicts $Co_3O_4$ to be a semiconductor, but severely underestimates the band gap. The PBE+U band gap (1.96 eV) agrees well with the available experimental value (~ 1.6 eV), whereas the band gap obtained using the PBE0 hybrid functional (3.42 eV) is strongly overestimated with respect to the available experimental data. While there is in principle no reason why the PBE0 functional should give a band gap in agreement with the experiment, the present result for $Co_3O_4$ is somewhat disappointing, as this functional has been found to yield satisfactory band gaps in several cases.[44,45]

MLWFs were used also to investigate the bonding properties of $Co_3O_4$. Independent of the electronic structure approach, we found 7 and 6 singly occupied d-type Wannier functions whose centers are very close to each cobalt ion at a tetrahedral and octahedral site, respectively. This is a clear indication that these ions have $Co^{2+}$ and $Co^{3+}$ oxidation states, in agreement with the formal oxidation states derived from simple chemical arguments. Besides d-type MLWFs on the cobalt ions, there are also Wannier functions with the character of $sp^3d$ bonds between cobalt and oxygen ions. Such hybridized bonding states imply the presence of a covalent component in the primarily ionic bonding character of $Co_3O_4$.

In agreement with experiment and consistent with crystal field theory (Fig.2), the computed magnetic structure is characterized by 3 unpaired spins on the $Co^{2+}$ ions and a weak coupling between the atomic magnetic moments. Due to the inclusion of exchange and correlation interactions in our calculations, however, the spin energy level distribution derived from the average energies of the d-type MLWFs (Fig.7) is quite different from the one given by simple crystal field theory. Both the PBE+U method and the hybrid PBE0 functional yield values of the magnetic moments which agree well with the experiment. This is an indication that these approaches provide a satisfactory description of the cobalt localized d states and, more generally, of the ground state properties of $Co_3O_4$.

## Acknowledgement

We are pleased to thank M. H. Cohen, P. Sit, F. Zipoli and R. Car for useful discussions on Wannier functions. This work was supported by DoE-BES, Division of Materials Sciences and Engineering under Award DE-FG02-06ER-46344, and Division of Chemical Sciences, Geosciences and Biosciences under Award DE-FG02-05ER15702.We acknowldege use of the TIGRESS high performance computer center at Princeton University which is jointly supported by the Princeton Institute for Computational Science and Engineering and the Princeton University Office of Information Technology.